\documentclass[twoside]{dis09}
\usepackage[latin1]{inputenc}
\usepackage[dvips]{graphicx,epsfig,color}
\usepackage{wrapfig,rotating}
\usepackage{amssymb,amsmath,array}

\begin{document}
\title{Semi--inclusive and exclusive measurements with EIC: \\
The advantages of lower energies}

\author{T.~Horn$^1$, P.~Nadel--Turonski$^{1, 2}$, C.~Weiss$^1$
\vspace{.3cm}\\
1-Jefferson Lab, Newport News, VA 23606, USA \\
2-Catholic University of America, Washington, D.C. 20064, USA}

\maketitle

\begin{abstract}
Exploring the nucleon's sea quark and gluon structure is a prime 
objective of a future electron--ion collider (EIC). Many of the
key questions require accurate differential semi--inclusive 
(spin/flavor decomposition, orbital motion) and exclusive 
(spatial distributions of quarks/gluons) DIS measurements in 
the region $0.01 < x < 0.3$ and 
$Q^2 \sim \textrm{few} \times 10 \, \textrm{GeV}^2$. 
Such measurements could ideally be performed
with a high--luminosity collider of moderate CM energy,
$s \sim 10^3 \, \textrm{GeV}^2$, and relatively symmetric configuration,
\textit{e.g.}\ $E_e/E_p$ = 5/30--60 GeV. Specific examples are
presented, showing the advantages of this setup (angular/energy 
distribution of final--state particles, large--$x$ coverage) compared
to typical high--energy colliders.
\end{abstract}

\begin{figure}[t]
\includegraphics[width=.99\textwidth]{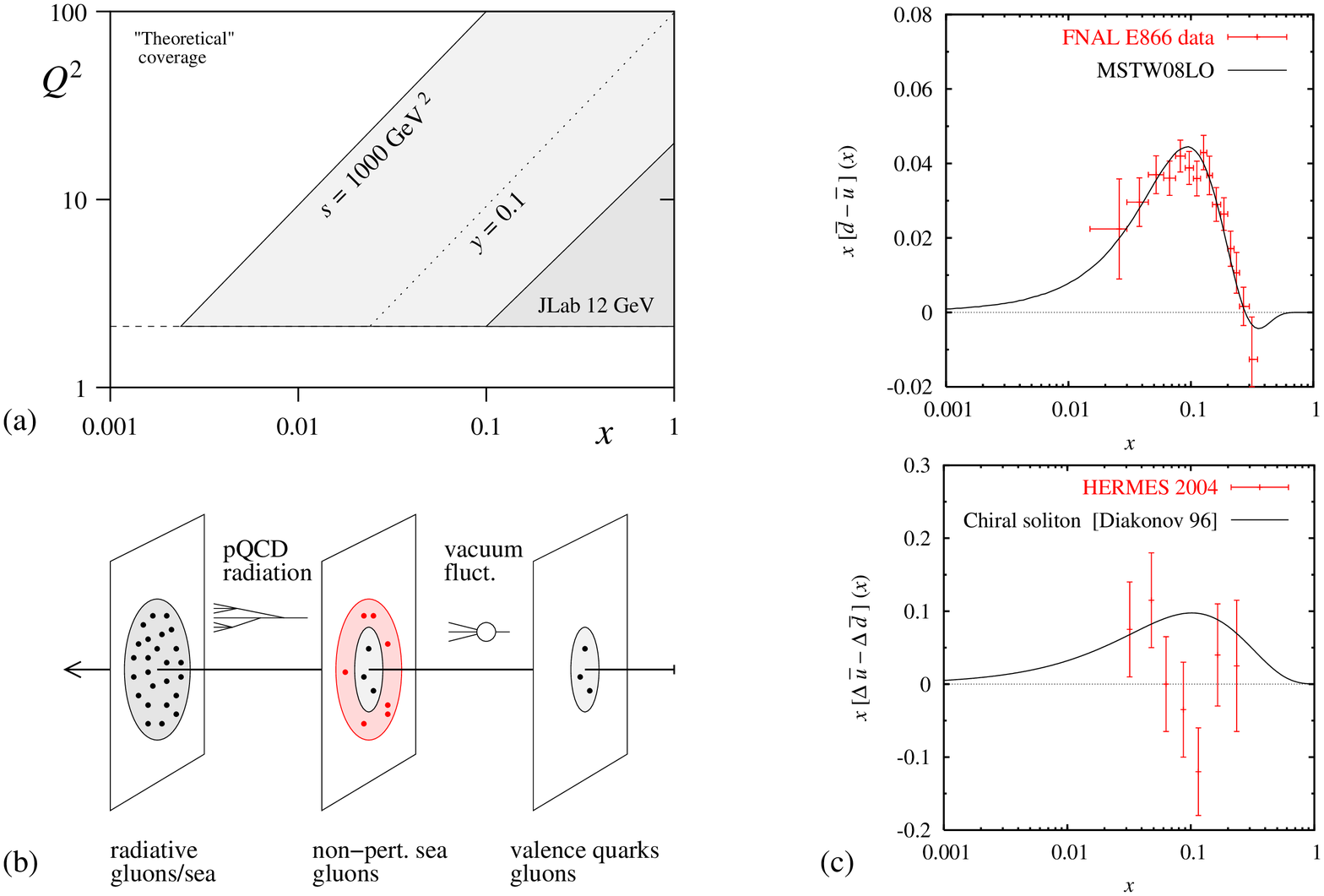}
\caption[]{(a) Kinematic coverage for DIS measurements with JLab 12 GeV
and a medium--energy EIC ($s = 1000 \, \textrm{GeV}^2$). (b) Components
of the nucleon wave function probed in DIS experiments at different $x$.
(c) Examples of non-singlet sea quark distributions.
Top: Flavor asymmetry $x \left[ \bar d - \bar u \right]$.
Bottom: Polarized flavor asymmetry $x \left[ \Delta \bar u - 
\Delta \bar d \right]$.} 
\label{fig:kinplane}
\end{figure}

One of the most fascinating aspects of the partonic description
of nucleon structure at short distances is that the nucleon becomes 
a many--body system, whose wave function has components with very
different number of particles. Depending on the excitation energy 
(or the Bjorken variable, $x$) DIS experiments can map the particle 
distributions in these different components, and thus provide much 
interesting information about the effective dynamics governing these 
degrees of freedom (see Fig.~\ref{fig:kinplane}a and b). Measurements at 
$x > 0.1$ probe mostly the valence quark component of the nucleon, 
for which a dynamical description as a few--body system, in the spirit 
of nuclear physics, seems to be appropriate. At higher energies, $x < 0.1$, 
one observes the quarks and antiquarks in the ``sea'' which are created by 
non--perturbative QCD vacuum fluctuations; their distributions carry 
quantum numbers (spin/flavor) and depend in a delicate way on the 
character of these vacuum fluctuations and their coupling to the 
valence quarks. Some of these $q\bar q$ pairs in the sea develop into 
the nucleon's pion cloud, which makes a distinctive contribution to 
the partonic structure at transverse distances $\sim 1/M_\pi$ and
is governed by chiral dynamics. Also in this region of $x$, gluons become 
an essential part of the nucleon's structure. At even higher energies, 
$x \ll 0.01$, the relevant degrees of freedom are radiatively 
generated gluons and singlet quarks, and the main dynamical questions 
concern the dominant characteristics of the radiation processes 
and the role of unitarity in the regime of high parton densities.
%
%

Mapping the nucleon's valence quark distributions is the objective
of the Jefferson Lab 12 GeV Upgrade; in particular, it will allow
one to access the unexplored region of $x \rightarrow 1$, and to map 
the transverse spatial distribution of the valence quarks
(generalized parton distributions, or GPDs) through high--$Q^2$ exclusive 
processes \cite{12GeV_CDR}. Exploring the nucleon's sea quark and gluon
structure will be the domain of a future EIC. Most of the interesting
non--singlet sea quark distributions, which are largely unaffected 
by perturbative QCD radiation (evolution) and directly testify to
the non--perturbative structure of the nucleon and the QCD vacuum,
are localized in the region of moderately small $x$, $0.01 < x < 0.3$. 
As an example, Fig.~\ref{fig:kinplane}c shows the 
antiquark flavor asymmetry $x \left[ \bar d - \bar u \right](x)$ 
measured in Drell--Yan pair production, as well as its polarization
predicted by a dynamical model of nucleon structure. Exploration of
such details of the sea quarks' spin and flavor distributions
is possible with semi--inclusive DIS measurements. Equally fundamental 
are their transverse spatial distributions, which are probed in 
exclusive processes. Gluons at $x > 0.01$, including their substantial
density in the region $x > 0.1$, are another essential --- 
and largely unexplored --- part of nucleon structure. They are probed in
semi--inclusive production of heavy quarks (open charm \text{etc.}),
and their spatial distributions are revealed in exclusive $J/\psi$
production. 

Semi--inclusive and exclusive measurements require high luminosity 
(low rates at high $p_T$ / high $Q^2$, several kinematic dependences) 
and the capability to analyze the final state 
(exclusivity, particle ID, momentum resolution).
Furthermore, in order to test the partonic reaction mechanism 
and extract information about nucleon structure it is essential to
perform fully differential measurements, in which the kinematic
dependences on $x$ and the final--state variables ($z$ and $p_T$ 
in semi--inclusive, $t$ in exclusive DIS) are measured at fixed $Q^2$, 
without correlations between the variables. Realizing such measurements with 
an EIC poses a major challenge for the accelerator as well 
as detector and interaction region design.

The EIC designs discussed so far have mostly focused on high CM energies
(eRHIC: $E_e/E_p =$ 10/100--250 GeV \cite{eRHIC_ZDR}, 
ELIC: $E_e/E_p =$ 3--7/30--150 GeV \cite{Bogacz:2007zza}), 
driven primarily by the desire to reach 
the saturation regime in $eA$ collisions, and also to extract the 
polarized gluon density from the $Q^2$--dependence of inclusive DIS data.
Similar to HERA, the proton energy is 10--20 larger than the electron 
energy in these collider designs. While the region of 
interest for sea quarks in nucleon structure, 
$0.01 < x < 0.3$ and $Q^2 \sim \textrm{few} \times 
10 \, \textrm{GeV}^2$, is formally within the kinematic coverage, 
significant restrictions for semi--inclusive and exclusive 
measurements appear when taking into account 
detector coverage and resolution. Recent studies show 
that such measurements can be performed much more efficiently with a 
collider of lower, more symmetric energies. In this note we summarize two 
of the pertinent considerations: (a) $x$--$Q^2$ coverage and final--state 
particle energies in semi--inclusive DIS; (b) angular distributions and 
$t$--resolution in exclusive processes. The arguments presented here
are general and do not assume a specific detector; they are intended 
to provide guidance for the detector design and eventual more 
detailed simulations.

{\bf Semi--inclusive DIS.}
Semi--inclusive DIS is an essential tool for the flavor decomposition 
of the nucleon's polarized valence and sea quark distributions. 
It is also used to study quark orbital motion and QCD final--state 
interactions through $p_T$--dependent observables (single--spin 
asymmetries, \textit{etc.}) which sit mostly at $x > 0.1$.
Both types of studies require fully differential measurements in 
$x, Q^2, z$ (and $p_T$) in the region $0.01 < x < 0.3$.
The ability to measure kinematic dependences at fixed $Q^2$ is
particularly important for testing the partonic reaction mechanism 
(separating leading and higher twist), and to make contact with the 
fixed--target results for these observables (JLab 12 GeV).

%
%
\begin{figure}
\parbox[c]{.5\textwidth}{
\includegraphics[width=.5\textwidth]{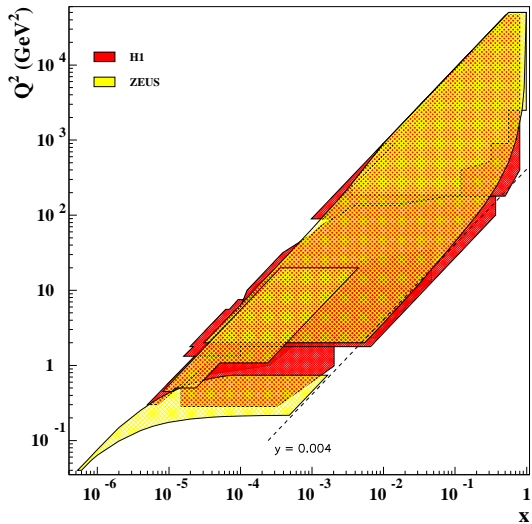}\vspace{-1ex}}
\hspace{.05\textwidth}
\parbox[c]{.4\textwidth}{
\caption[]{Kinematic coverage in DIS experiments at HERA.
Coverage at moderately small $x$ ($> 10^{-2})$ 
and fixed $Q^2$ is limited by event reconstruction. 
The electron method is typically used for $y > 0.1$; 
hadronic methods have achieved $y_{\rm min} \approx 0.005$ 
in inclusive DIS. (We thank P.~Newman for the computer
code used to generate this drawing.)} 
\label{fig:xq2plane}
}
\end{figure}
In a high--energy collider, the kinematic reach for DIS measurements
at large $x$ is limited by the experimental limits on the
$y$ variable, $y = Q^2/(x s) > y_{\rm min}$, which implies
that large values of $x$ are accessible only at high $Q^2$. 
If the electron variables are used for event reconstruction the 
resolutions in $y$ and $x$ diverge for $y \rightarrow 0$; 
using methods with hadronic variables \cite{Bassler:1997tv} 
$y_{\rm min} \approx 0.005$ was reached in inclusive measurements 
at HERA (see Fig.~\ref{fig:xq2plane}), albeit with considerable uncertainties.
A medium--energy collider, providing high luminosity over a range of 
$s$ around $\sim 1000 \, \textrm{GeV}^2$ (\textit{e.g.}, 
$E_e/E_p =$ 3--7/30--60 GeV) could cover the region $0.01 < x < 0.3$ 
at $Q^2 \sim$ few $\times 10$ GeV$^2$ with comfortable values of $y$ 
(see Fig.~\ref{fig:kinplane}a), enabling precise and fully 
differential SIDIS measurements at fixed $Q^2$.
Studies show that \textit{e.g.}\ the flavor separation of 
polarized quark distributions
at $x > 0.01$ can be performed much more efficiently at 
$s \sim 10^3 \, \textrm{GeV}^2$ than at $10^4 \, \textrm{GeV}^2$, 
as the lower CM energy permits measurements at lower $Q^2$ where the 
cross sections are larger.

Another important issue in SIDIS is particle identification (PID). 
An advantage of the lower CM energy is that final--state particles 
are produced at lower energies, so that the range where good PID
is available, $E_h =$ 1--10 GeV, matches the
$z$--range of physical interest, $z > 0.1$. Energy resolution is
likewise improved.

{\bf Exclusive processes.}
Exclusive processes $eN \rightarrow e' M N \; (M = \textrm{meson}, \gamma,
\ldots)$ at $Q^2 \sim \textrm{few} \times 10 \, \textrm{GeV}^2$
reveal the transverse spatial distribution
of quarks and gluons in the nucleon and its change with $x$ 
(``nucleon tomography''). It is encoded in the $t$--dependence 
of the GPDs, which can be extracted from that of the differential 
cross sections.
In order to resolve interesting details of the quark/gluon spatial 
distribution, such as the contribution of the pion cloud,
accurate measurements in the range $|t| \ll 1 \, \text{GeV}^2$
are needed. Measurements of exclusive processes are among the most demanding
applications of a future EIC. Besides the need for high luminosity
(small cross sections, fully differential measurements), 
the challenges are to ensure exclusivity 
of the events and to achieve high resolution in $t$ at small $|t|$. 
Both are greatly aided by choosing collider energies appropriate to 
the $x$--range in question.

\begin{figure}[t]
\hspace{.08\textwidth}
\includegraphics[width=.80\textwidth]{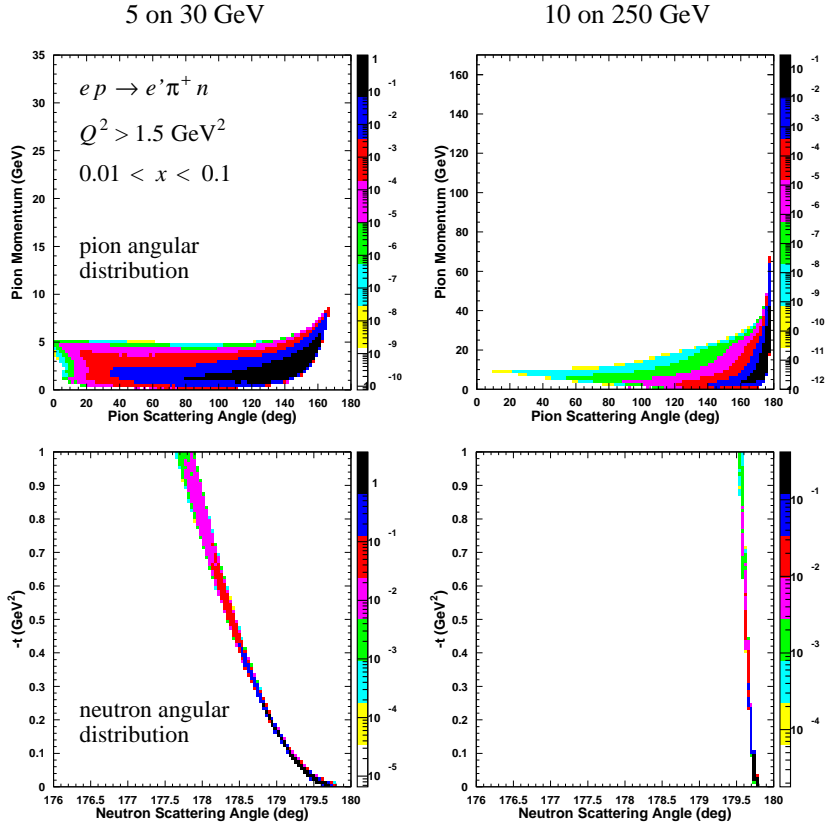}
\caption[]{Combined angular and momentum distribution of the $\pi^+$ 
(top row), and angular and $|t|$--distribution of the recoil $n$ 
(bottom row), in exclusive production $ep \rightarrow e'\pi^+ n$
with two different $ep$ colliders: $E_e/E_p =$ 5/30 GeV
(left column), and 10/250 GeV (right column). A lab angle
of $0^\circ$ corresponds to the electron, $180^\circ$ to the 
proton direction. Shown are the results of simulations based on 
a specific model of the exclusive cross section.} 
\label{fig:theta}
\end{figure}

As an example, Fig.~\ref{fig:theta} shows a simulation of exclusive
$\pi^+$ production, $ep \rightarrow e' \pi^+ n$ in DIS kinematics, 
which provides interesting information about the nucleon's partonic 
structure in the region $x > 0.01$ (non-diffractive process) 
and about the pion form factor. It compares the angular distributions 
of the produced pion and the recoiling neutron for an $ep$ collider
with $E_e/E_p =$ 5/30 GeV and 10/250 GeV.
With the lower--energy collider the pions are distributed over
a wide angular range and could be detected with 
a central detector; with the high--energy collider they are kinematically
boosted forward and appear under very small angles. 
Similarly, with the lower--energy collider 
the recoiling neutron emerges at larger angles (several degrees),
allowing for much better $t$--resolution with an appropriate
forward detector. The relation between $|t|$ and $\theta_n$ 
at small angles is $|t| \approx E_p^2 (\theta_n - \pi)^2$, {\it i.e.},
at the same CM energy, a more symmetric collider with lower proton 
energy offers much better prospects for good $|t|$--resolution.
Similar conclusions apply to other meson production processes,
\textit{e.g.}\ exclusive $J/\psi$ production, which maps the
transverse spatial distribution of gluons in the nucleon,
as well as to deeply--virtual Compton scattering.
%
%

{\bf A high--luminosity medium--energy EIC for nucleon structure.}
A design for a high--luminosity medium--energy EIC for nucleon 
structure measurements is being developed at JLab \cite{Thomas:2009ei}.
This ring--ring collider, which uses the upgraded CEBAF
complex for electron acceleration, would operate with
energies in the range $E_e/E_p =$ 3--11/12--60 GeV and 
can achieve luminosities of up to 
$\textrm{few} \times 10^{34} \, \textrm{cm}^{-2} \textrm{s}^{-1}$ 
over a broad range of CM energies,
$s =$ few 100--2600 $\textrm{GeV}^2$. The envisaged
proton/ion complex uses an SRF linac and a pre--booster ring for 
acceleration and cooling. The figure--8 design of the electron
and proton/ion storage rings 
(circumference $\sim 600 \, \textrm{m}$) guarantees optimal beam 
polarization and allows for up to 4 interaction regions.
An upgrade to the high--energy ELIC ($s \sim 10^4 \, \text{GeV}^2,
L \sim 10^{35} \, \textrm{cm}^{-2} \textrm{s}^{-1}$)
would be possible with larger rings. The medium--energy EIC
would offer unprecedented capabilities for exploring the nucleon's
sea quark and gluon structure through semi--inclusive and exclusive 
measurements, and for studying the structure of nuclei with novel 
short--distance probes. It supports a physics program distinct from 
that of the high--energy stage, and would thus add considerably
to the overall potential of an EIC.

This note reports about results of collaborative work 
and discussions with A.~Bogacz, Ya.~Derbenev, R.~Ent, Ch.~Hyde, 
G.~Krafft, A.~W.~Thomas and Y.~Zhang.

\begin{footnotesize}
Notice: Authored by Jefferson Science Associates, LLC under U.S.\ DOE
Contract No.~DE-AC05-06OR23177. The U.S.\ Government retains a
non--exclusive, paid--up, irrevocable, world--wide license to publish or
reproduce this manuscript for U.S.\ Government purposes.
\end{footnotesize}

\begin{footnotesize}
\end{footnotesize}
\end{document}